\documentclass[prb]{revtex4}
\usepackage{epsf}
\begin{document}
\title{COMETS,  HISTORICAL  RECORDS  AND VEDIC LITERATURE}
\author{Patrick Das Gupta}
\affiliation{Department of Physics and Astrophysics, University of Delhi, Delhi - 110 007 (India)}
\email{pdasgupta@physics.du.ac.in}

\begin{abstract}
  A verse in book one of  Rigveda  mentions  a cosmic  tree  with rope-like aerial roots held  up in the sky.  Such an imagery might have ensued from the appearance of  a comet having `tree stem' like tail, with  branched out portions resembling aerial roots.  Interestingly enough, a comet referred to as `heavenly tree' was seen in 162 BC, as reported by old Chinese records.  Because of weak surface gravity, cometary appendages   may possibly  assume strange shapes depending on factors like rotation, structure and composition of the comet as well as solar wind pattern. Varahamihira and Ballala Sena  listed several comets having strange forms as reported originally  by ancient seers such as Parashara, Vriddha Garga, Narada  and Garga.

Mahabharata  speaks  of  a mortal king Nahusha  who ruled the heavens when   Indra,  king of gods,  went into hiding.  Nahusha became  luminous and egoistic     after absorbing radiance from gods and seers. When he kicked    Agastya (southern star Canopus), the latter cursed him to  become a serpent and fall from the sky. We posit   arguments to surmise that this  Mahabharata lore is a mythical recounting of a cometary event wherein a    comet crossed Ursa Major,  moved southwards with an elongated tail in the direction of  Canopus and eventually went out of sight.  In order to check whether such a conjecture is feasible, a preliminary list of comets (that could have or did come close to Canopus) drawn from various historical records   is presented and discussed.

\end{abstract}
\maketitle
\section{Introduction}
Stars and constellations were observed with keen interest by Indo-Aryans, mainly for  ascertaining   auspicious times to  perform various  vedic rituals and  sacrifices  by frequently  monitoring  apparent stellar motions (Pingree, 1981).  It is, therefore, inconceivable  that  comets of antiquity visible to naked-eye would have escaped  attention of vedic seers. After all,   bright comets approach the Sun roughly once every five years (Hasegawa, 1980; Ho 1962).   However,   studies on ancient comets  in traditional astronomy of India is handicapped by the fact  that, unlike the existing  systematic  astronomical  records of  the Greeks and the Chinese describing pre-telescope era comets (Fotheringham, 1919; Ho, 1962;  Pingre, 1783; Williams, 2014), ancient Sanskrit  texts  on  cometary events  have been lost in the ravages of time. 

Because of   the   absence   of   older  historical records, cometary references often get mingled with myths and  lore,  making it difficult to establish firmly any  inference on  comets of antiquity in the Indian context.  Varahamihira's Brhat Samhita (BS) of 550 AD cite many older literature on comets  attributed to ancient seers like Parashara,  Garga, Vriddha Garga  and Narada, that are no longer extant (Bhat, 1981). In order to salvage some accounts of comets that might have been sighted from India in ancient   times,  one  is forced to fall back on commentaries   provided in BS (Kochhar, 2010) and in Ballala Sena's   Adbhuta Sagara (Iyengar, 2008; 2010). 

 Halley's own work shows that study of historical records of comets is very important for astronomy. This is evident from   extant medieval cometary records of 1337-1698 CE that enabled Halley to characterize orbits of 24 comets (Halley, 1705). This was  followed  up with  estimation of  orbital  parameters of various comets using data gleaned from old Chinese and European sources (Hasegawa, 1979; Hasegawa, 1980; Ho, 1962; Kiang, 1972; Pingre, 1783). Comets and novae observed by  Koreans in the bygone era were collected by Sekiguchi (1917a,b).  Similarly, Kanda (1935, 1947) put together  old Japanese records of astronomical significance and published them.  A large collection of  Chinese records, along with Japanese and Korean ones, of  past comets and novae were compiled  and  translated  into English by Ho (1962).
 Kiang (1972) made use of these records to compute  orbital trajectory of Halley's comet   over  past 28 revolutions,  going all the way back  to 239 BC. Later,  Yeomans and Kiang (1981) retraced the comet's trajectory to 1404 BC by numerically integrating the equations of motion. In an exciting development, evidence of  Halley's comet sightings in 164 BC and in 87 BC were  discovered   in   Babylonian clay tablets (Stephenson, Yau  and Hunger, 1985) referred  to as `regular watchings' or Diaries,  containing daily observations of the sky that were likely to have been commissioned by priests of Marduk temple  (Brown, 2002; Sachs, 1974). Several ancient Assyrian and Babylonian scholars had commented  on a comet that appeared late in 675 BC which is yet to be identified (Brown, 2002).

In what follows, a vignette about comets, in the context of vedic literature, is provided  in section 2.  Section 3 gives a brief account of the Nahusha lore and  surmises  that this is a mythical retelling of a cometary event. Arguments are put forward  in support of the conjecture, shortly thereafter. A preliminary study is undertaken in section 4 mainly to shortlist  and discuss comets described in historical records that either appeared near Canopus or could have been seen near it.   Finally, in section 5, we conclude by reflecting on the  possible allusions to comets in vedic texts. 

\section{DHUMAKETU AND  KETU}
The Sanskrit word dhumaketu literally means `smoke banner' and it  appears in  about  half a dozen hymns of Rigveda,  while  Atharvaveda has  a hymn about  Saptarishi  (Ursa Major) being veiled by a dhumaketu, entailing   that this word  connoted a comet (Iyengar, 2010) more than 3000 years ago.  It is to be noted that the mantra portions of Atharvaveda have been dated to about 1150 BC   because of their   first direct mention of  metal iron  (Witzel, 1995).  Ketus (in plural),   meaning   rays of light or fire-smoke combine, have also been discussed in Atharvaveda, and it is very likely that  they  represented comets or meteors (Kochhar, 2010).

Book one of Rigveda has two hymns (1,24,7) and (1,24,8) which have been translated  by Griffith (1896, p.14) as follows:
\vskip 1 em
7. Varuna, King, of hallowed might, sustaineth erect the Tree's stem in the baseless region.
Its rays, whose root is high above, stream downward. Deep may they sink within us, and be hidden.
\vskip 1 em
8. King Varuna hath made a spacious pathway, a pathway for the Sun wherein to travel.
Where no way was he made him set his footstep, and warned afar whate'er afflicts the spirit.

According to Parpola (2009, 2010),  these verses were composed around 1000 BC, and could be interpreted as god Varuna, the guardian of cosmic law, holding up the aerial roots of a cosmic banyan tree  in the sky. Imagery of a tree stem up in the sky, with long aerial roots, suggests that the simile  was inspired by an apparition of a comet of bygone era possessing a `tree stem' like tail with branched out portions resembling rope-like aerial roots (Das Gupta, 2015).  It is noteworthy to point out that Chinese records speak of  tianchan (`heavenly tree'),  a comet appearing in the southwest on the evening of February 6 in 162 BC (Ho, 1962). Owing to weak surface gravity, cometary appendages resulting from interaction with solar wind and radiation pressure may assume strange shapes depending on spin, structure and composition of   the  comet as well as solar wind pattern.

Later  texts belonging to Puranas  (containing old Hindu royal genealogies as well as mythologies) explain the phenomena of  stars and planets  going around a fixed Dhruva (Pole star), instead of falling down,  by  invoking  invisible  rope like aerial roots growing outwards from the Dhruva and attaching themselves to   celestial objects  (Parpola, 2010). Again, one may speculate that this imagery owes its origin to a comet of antiquity that had appeared in the vicinity of the Pole star.

Varamihira's BS, after delineating  features of 1000 comets,  states that it is impossible to determine the rising and setting of  comets from any kind of calculation (Subbarayappa and Sarma,1985). Motion of a comet named Chala Ketu   (literally, moving comet) was  described vividly by Varahamihira,  emphasizing  its rise on the west and  increase in the tail size  as it proceeded  towards north,  and eventually making   contact with Ursa Major (Chandel and Sharma, 1991).  Anticipating  periodic paths, the seer Narada had stated – `there is only one comet which comes time and again', while  Bhadrabahu  estimated  comets to be   hundreds in number, each with different period (Sharma, 1986; Chandel and Sharma, 1991).

Parashara (who was likely to have lived around 1000 - 700 BC)  had cataloged 101 comets, 26 of which (that included Chala Ketu) were  described in great detail (Iyengar, 2006).  Seer Garga, belonging to about 100 BC (Kane, 1975; Kochhar, 2010),  had listed 77 comets that were characterized by a dark reddish hue, as cited in  BS (Bhat, 1981;  Kochhar, 2010).  It is possible that a descendant of   Garga had composed Gargya jyotisha between 1 BC and 1 AD, in which  both Rahu and Ketu are included in the list of nine grahas (i.e. `grabbers' or planets in the Indian context), with Ketu representing comets, and not the dismembered torso of Rahu (Yano, 2003).  One may recall that Rahu and Ketu (as a proper name)  are associated with eclipses (Kochhar, 2010). 

In Puranic texts, Rahu is a demon who partook  celestial ambrosia in a clandestine way to attain immortality.  To punish him,  Lord Vishnu  hurled his discus  at Rahu and severed  the head from the torso. The tail like torso  was christened Ketu, most likely  because  ketus (comets) generically have tails (Das Gupta, 2015).  Speaking of  Lord Vishnu, it is interesting to note that Jayadeva (the 12-th century temple poet of Puri Jagannath temple) had described  Kalki, the last avatara of  Vishnu,  carrying a scimitar that blazes like a comet. Could he have been influenced by Ballala Sena's commentaries on comets?   Ballala Sena was a king of Mithila and Vanga (not far from Puri) who had written a treatise named `Adbhuta Sagara' (Ocean of Wonders) sometime around 1100-1200 AD. Adbhuta Sagara described collections of comets   originally due to  seers Parashara, Vriddha Garga, Garga, Atharva, Varahamihira and  Asitadevala (Iyengar, 2008; 2010).

Several Hindu temples  have  sculptures on  the lintel of their entrance doors  representing  nava grahas (nine `planets') with Ketu  depicted as having an anthropomorphic bust along with a serpentine tail.   Atharvaveda-Parishishtha  contains  verses  not only  about grahas, nakshatras (lunar mansions), and rahu but also about  ketus  (comets) classified  according to seasons (Miki and Yano, 2010). However, many of its chapters  were composed after Greek astrology was introduced in India around 300 AD.

\section{CANOPUS, URSA MAJOR  AND THE  NAHUSHA MYTH}
Canopus or Alpha Carinae is a -0.73 magnitude, spectral F0 type supergiant, which is about 200000 times more luminous than Sun, and  is located 60-80 pc away from us (Achmad, De Jager and Nieuwenhuijzen, 1991). In the Northern Hemisphere,  Canopus (also called Agastya in India) is visible during the winter season from regions south of 37 degree latitude.  Agastya was  a seer who composed around 27 hymns of Rigveda (Mahadevan, 1986).  There is also a hymn in Rigveda likely to be due to Agastya in which Pleiades is mentioned (Das Gupta, 2015). In  India, Alpha Carinae  has been associated with the name Agastya  since about  600 BC (Ghurye, 1977, p.123-125). According to the Puranic literature, he was the first vedic Aryan to cross the Vindhya hills to explore  the southern regions of India (Abhyankar, 2005). 

 It is interesting to note that the  older name of Canopus is Alpha Argus as it is associated  with the southern constellation Argo Navis. Although this name sounds like Argha, a celestial ship steered by Agastya so that  the Sun  could sail across the sky  (Allen, 1963, p.66),  it originated from  the myth of Jason and the Argonauts.  In Greek, argo means swift and Argo Navis stands for the swift ship  that was built by Argos for Jason. At this point, it is pertinent to point out that Arka,  a Sanskrit word phonetically similar to  Argha, is also a name of the Sun, with arka  connoting  `a ray of light' in Sanskrit, much like the word ketu. 

In India, Ursa Major is called Saptarishi (i.e.  seven seers),  rishi being a Sanskrit word for  sage or seer.   However, older vedic literature refers to  the  Big Bear  as `rikshas', which means  bears  in archaic Sanskrit.  This clearly indicates a common Indo-European origin of the vedic people, as the older connotation of Big Bear for the constellation survived  till about 1000 BC or earlier (Ghurye, 1972, p.102-103). Later,  `rikshas' might have got substituted by the similar sounding `rishis'  (which has a different meaning altogether) as far as this constellation  was concerned. (Stars of the Ursa Major had been identified with the seven seers of Rigveda by about 900 BC (Ghurye, 1972, p.114-120)).

Mahabharata, which is often referred to as the fifth veda,  narrates the strange story of Nahusha, a human king,  who took charge  as the king of  gods  when Indra went incognito after killing his arch foe, Vrtra.  According to Hiltebeitel (1977, p.332),  Nahusha then turned radiant with `five hundred lights on his forehead burning' as he drew energy from  seers, demons, gods, goblins, etc., and reigned over the sky. In order to seek attention of  Indra's consort Sachi, he forced the seven seers of Ursa Major to carry him around in a palanquin. Seer Bhrigu, one of the seven sages, seethed with rage because of this humiliation he was subjected to.   He requested Agastya to temporarily substitute him and lend his shoulder to the  carriage. As Agastya was quite  short in height, the palanquin with Nahusha in it, lost its balance when he took Bhrigu's place as a bearer. Tilting of the carriage infuriated  Nahusha so much that he angrily kicked Agastya.  An enraged Agastya thereby cursed the king to turn into a serpent and fall from the sky (Hiltebeitel, 1977).

There are several features in the above lore  that lead one to speculate that a very old cometary event, in which the tailed visitor trespassed  Saptarishi constellation from north with its tail gradually increasing in size as it moved southward towards the  star Agastya, and eventually went out of sight as it dipped below the horizon,  metamorphosed into a mythical story  (Das Gupta, 2015).  Let us go through the key points one by one.

1.	Nahusha is intimately linked with celestial objects since he is a son of the daughter of Svarbhanu, the eclipse causing demon of Rigveda (Griffith, 1896; Kochhar, 2010; Vahia  and Subbarayappa, 2011), who later  got associated with Rahu and Ketu (Kochhar, 2010).  

2.	He also belongs to the lunar dynasty with ancestors such as Moon,  
 Mercury and   Atri,  who is one of the seers/stars of Ursa Major (Hiltebeitel 1977).

3.	Big Bear is also referred  to as cart or  `wain' (Ghurye, 1972; Hiltebeitel, 1977) and hence could have become a mythical `carriage' or a `palanquin'  over the years when the comet sighting story was being propagated.

4. Varahamihira had mentioned (as paraphrased by Al Biruni)`Comets are such beings as have been on accounts of their merits  raised to heaven, whose period of dwelling in heaven has elapsed and who are then redescending to the earth' (Allen, 1963, p.27).

5.	Varahamihira  had prescribed worship of Agastya for kings, and had stated categorically that if this southern star  is struck by a comet or a meteor there would be famine (verse 22 of BS;  Bhat, 1981). An important question is:  was he aware of the Nahusha lore?

6.	According to many Puranic texts,  the eclipse causing demon, Rahu, had a serpentine form with just a head and a tail (Kochhar, 2010), very much like a comet. After the demon had surreptitiously tasted the ambrosia that led to immortality, Rahu was struck by Vishnu's discus as an act of retribution (Das Gupta, 2015). Its severed  tail was christened as Ketu, a proper noun inspired by the common noun ketu that represents a tailed comet.  Hence, it is not a far fetched  idea to associate a comet  with Nahusha turning into a serpent.

In short,  narration of the event  wherein a comet traversed across the Big Bear  could have created  an  imagery in the listener's mind in which a radiant object was initially carried by Saptarishi (cart or  `wain' ). Then the  recounting of comet's motion southwards  with its tail growing longer, and eventually making  an apparent contact with Canopus before going out of sight,  could have conjured up an image of  Nahusha kicking Agastya and  disappearing from the sky thereafter.

\section{HISTORICAL COMETS AND CANOPUS}

An interesting exercise that could be undertaken is to study  comets that were observed in the vicinity of Canopus or those which could have been near the southern star so that one may attempt  constraining  the data set keeping in mind the conjecture of Nahusha myth being a retelling of a past cometary event.  In the northern hemisphere, Canopus can clearly be spotted below Sirius during  the months of December to March, south of 37 degree latitude. 

 To make  a beginning,  we  look at reports of comets that came very near  the southern star. We also make a preliminary study of   far eastern historical records  of  ancient comets and, in particular,  Halley's comet sighted between 240 BC and 530 AD.  Although   Lao-jen,  meaning `The longevity star'  (Canopus), does not seem to be associated with any of the returns of Halley's comet in these records (Stephenson and Yau, 1985), it may still be a worthwhile exercise to list those  with perihelion passage times falling during the winter in the period between 240 BC  and 530 AD.  As it is unlikely that the Nahusha myth was added to Mahabharata post-Varahamihira, we consider its return only up to 530 AD.

A preliminary list of  observed  comets that came or could have come close to  Canopus has been provided below:
\vskip 2 em
1. While he was looking for Encke's comet from Italy in May, 1822, J. L. Pons chanced upon Comet C/1822 K1  (Kronk, 2003). This comet  moved  fast  southward, and could not be viewed from the Northern Hemisphere thereafter. It was spotted later near Canopus in June, 1822, from Rio de Janeiro.   Comet C/1822 K1 was  sighted about 3 degree away from Canopus on June 18, but by June 19, 1822, the angular separation between them had increased to about 12 degree  (Robertson, 1831). This comet has been classified as a hyperbolic comet since  it escaped the solar system  after making use of the Sun's gravity  to gain speed by virtue of  the  slingshot effect,  never to appear again.
\vskip 2 em

2. Comet C/1853 G1 was discovered by K. G. Schweizer on April 5, 1853, south of rho aquilae 
 and  which later showed up in the southern hemisphere on April 30, 1853, with its tail pointing towards Canopus. The tail  grew from about  4 degree to 8 degree   in length within a day and was seen  on June 11, 1853  (Kronk, 2003). The estimated period of  C/1853 G1  is about 782 years  and,  therefore, it could have been seen in  493 BC and in  289 AD (Das Gupta, 2015).
	
\vskip 2 em

3.	Catalogs  due to Ho (1962) and Xu, Pankenier and Jiang (2000) list a very large number of comets that appeared in the winter (of northern hemisphere).  Of course, one is aware of the caveat that report of  winter apparition does not mean that perihelion passage time fell in winter. In these records, comets are referred to either as stars becoming fuzzy or broom stars or extended vapour or guest star, etc. Lack of space restricts us to list  only those winter comets that appeared in the interval from 974 BC to 133 AD: 
\vskip 2 em
(a)	A star became fuzzy during February - April in 974 BC

(b)	A broom star appeared in the winter of 525 BC

(c)	During October - December months  stars became fuzzy in 482 BC and in 481 BC

(d)	In 238 BC, a broom star appeared in north and moved  southwards for 80 days

(e)	A broom star appeared in the east during February 19 - March 20 in 234 BC

(f)	`Heavenly tree' or a comet   appeared in the southwest  on February 6, 162 BC

(g)	A star in the southwest  became fuzzy during January 18 to February 16 in 154 BC

(h)	Between October 12 and November 10, a star became fuzzy in 147 BC

(i)	A star became fuzzy during February - April  in 120 BC

(j)	During February – April in 119 BC a star turned fuzzy

(k)	In 69 BC, a star in the west became fuzzy during January 27 - February 24

(l)	A star turned fuzzy in 32 BC between  February 6 and March 7

(m)	During January 10 – February 7, a streak of white vapour appeared in the southwest extending from the ground to the sky  in 5 BC
 
(n)	In AD 22, between November 13 – December 12, a star became fuzzy and moved southeast

(o)	During December 17, 46 AD - January 15, 47 AD, a fuzzy star appeared in south

(p)	Between December 6,  55 AD, and April 6, 56 AD, a comet appeared traveling southwestwards

(q)	In January, 78 AD, a star became fuzzy

(r)	A comet appeared in January, 101 AD

(s)	During January 9 - February 6, 110 AD, a broom star appeared in south

(t)	A guest star appeared on January 9, 117 AD, in the west

(u)	In 132 AD, a grayish star appeared on January 29 with vapours in the form of rays

(v) A comet with a long tail appeared southwest on February 8, 133 AD 

\vskip 2 em

4. Halley's comet:
\vskip 1 em
One may safely ignore apparitions of Halley's comet in 240 BC, 87 BC, 141 AD, 218 AD, 295 AD,  451 AD and 530  AD  since  the corresponding perihelion passage times fell in or after March but before October. That leaves only its appearances  in 164 BC, 12 BC, 66 AD and 374 AD for which the perihelion passage months were November, October, January and February, respectively (Hughes, 1985; Kiang, 1972;  Tsu, 1934; Yeomans, Rahe and Freitag,  1986). Past orbits of Halley’s comet have been well studied, and so it should not be very difficult to rule out possible proximity to Canopus during its  apparitions  in 164 BC, 12 BC, 66 AD and 374 AD.  

Of course, it is far from clear that any of the above comets listed from 3 to 4 came actually close to Canopus. However,  comets 3  (d), 3 (f),  3 (g) , 3 (m), 3 (o), 3 (p) and 3 (s) in the above list  appear promising as far as the possibility of their being seen near Canopus. It is interesting to note that the Chinese document of Se-ma Ts'ien mentions the apparition of  the Standard of Tch'e-yeou   in 134 BC, which was a comet that had a serpentine form in the shape of a standard (Chavannes, E., 1899).  Its appearance  had also been reported by Hipparchus (Fotheringham, 1919). Perhaps it  is  the same comet (no. 39) listed by Ho (1962) and by Xu, Pankenier and Jiang (2000)  that was seen  in the east during August 31 to September 29 in 135 BC, stretching across the  entire sky.

 Comet  (no.39) of 135 BC was not sighted in the winter as par  the historical records (Ho, 1962). However,  if it is identified with the Standard of Tch'e-yeou   then it is plausible that  it could have crossed the perihelion  in or after January and reappeared in 134 BC. Then, it makes sense to short list this comet too, particularly because of its serpentine shape. It is noteworthy to point out that, according to Fotheringham (1919),  this comet returned during 120-119 BC. In that case,  it could correspond to the winter apparitions 3 (i) and 3 (j), making it a comet with 15 years orbital period. According to Kochhar (2010), date of the closure of Mahabharata is likely to be 100 BC, in which case one may surmise that the apparition of Standard of Tch'e-yeou  in 134 BC could have given rise to the Nahusha myth.

\section{Conclusions}
While Halley's pioneering work of extracting information on  cometary orbits from medieval  records proved so useful to astronomy, paving the way for  further comet  research based on far eastern historical catalogs, one encounters a serious setback  in the Indian scenario since ancient Indian  records of comets are no longer extant.  As a consequence it is difficult to separate   real cometary references from  myths and  lore. Nevertheless, it is important to look for allusions in vedic texts  to  dhumaketu, ketus as well as strange  forms (e.g. serpentine or aerial root-like) in the sky (as vedic priests were enamoured by celestial objects, chiefly for time-keeping purposes) with the hope that something significant on comets turn up.  

Mention of a cosmic banyan tree, in a late Rigvedic verse, held up in the sky does entail one to speculate that it was  inspired by a comet of antiquity, particularly because there exists a reference in old Chinese records  to a 162 BC comet as the `heavenly tree'. Similarly, the myth in which Nahusha turned into a serpent after kicking Agastya and fell from the  sky leads one to surmise that description of  a comet that trespassed Ursa Major, moved southwards growing in length, crossed Canopus and went below the horizon turned gradually into a lore as it got passed around. The strongest arguments in favour of this interpretation comes from Varahamihira's (a) statement - `Comets are such beings as have been on accounts of their merits  raised to heaven, whose period of dwelling in heaven has elapsed and who are then redescending to the earth' (Allen, 1963, p.27) and (b) instruction that  kings must  worship Canopus (Agastya) and that if a comet strikes this southern star, there will be calamities (Bhat, 1981). 

There are 19-th century reports of comets that appeared  very close to Canopus. Although,  ancient far eastern records of comets do not directly mention comets near Lao-jen (Canopus), reference to comets of older times that either moved southwards or were seen in the south survive. Comet that looks very promising, as far as the Nahusha myth is concerned, is the Standard of Tch'e-yeou  which was sighted  in 134 BC to have a  peculiar  serpentine form.  If one takes 100 BC to be the epoch  of closure of Mahabharata then it is plausible that the Nahusha lore grew out of the apparition of this strange comet. Needless to point out that more work is required in this area and one must also study thoroughly the past trajectory of  Halley's comet to check whether it could have appeared very close to  Canopus.

\section{Acknowledgements}
 It is a pleasure to thank Professor Mayank Vahia for organizing an academically stimulating international conference on Oriental Astronomy  at  Pune in 2016, and for his constant encouragement to complete this article. I am indebted to Professor W. Orchiston for examining the article critically as well as providing invaluable suggestions. I also thank Dr. N. Rathnasree for helpful discussions on this subject.
  
\vskip 2.0 em
{\bf{References}}
\vskip 2.0 em
Abhyankar, K. D., 2005.  Folklore and Astronomy: Agastya a sage and a star. {\it Current Science}, 89, 2174-2176. 
\vskip 2.0 em
Achmad, L., De Jager, C. and Nieuwenhuijzen, H., 1991. Atmospheric model parameters and shock wave field  for the  supergiant Alpha Carinae (F0Ib). {\it Astronomy and Astrophysics}, 249, 192-198. 
\vskip 2.0 em
Allen, R.H., 1963. {\it Star names: their lore and meaning}.  New York, Dover.
\vskip 2.0 em
Bhat, R. M., 1981. {\it Varahamihira's Brhat Samhita}. Delhi, Motilal Banarsidass Publishers.
\vskip 2.0 em
Brown, D.R., 2002. Babylonian observations. {\it Highlights of Astronomy}, 12, 311-316.
\vskip 2.0 em
Chandel, N. K. and  Sharma, S., 1991. A Comparative Study on Cometary Records from the Brhat Samhita and Bhadrabahu Samhita. {\it Indian Journal of History of Science}, 26, 375-382.
\vskip 2.0 em
Chavannes, E., 1898. {\it Les Memoires historiques de Se-ma Ts’ien}, Volume 3. Paris, E. Leroux.
\vskip 2.0 em
Das Gupta, P.,  2016.  Comets in Indian Scriptures. In  Abbott, B. P. (ed.). {\it Inspiration of Astronomical Phenomena VIII: City of Stars}. ASP Conference Series,  Volume 501. San Francisco, Astronomical Society of Pacific. Pp.151-160.
\vskip 2.0 em
Fotheringham, J.K.,  1919. The new star of Hipparchus, and the dates of the birth and accession of Mithridates. {\it Monthly Notices of Royal Astronomical Society}, 79, 162-167.
\vskip 2.0 em
Ghurye, G. S., 1972. {\it Two Brahmanical Institutions - Gotra and Charana}. Bombay, Popular Prakashan.
\vskip 2.0 em
Ghurye, G. S., 1977. {\it Indian Acculturation: Agastya and Skanda}.  Bombay, Popular Prakashan.
\vskip 2.0 em
Griffith, R. T. H., 1896. {\it The Hymns of the Rgveda}.  Delhi, Reprinted by Motilal Banarasidass Publishers.
\vskip 2.0 em
Halley, E., 1705. {\it A Synopsis of the Astronomy of Comets}. London, printed for John Senex.
\vskip 2.0 em
Hasegawa, I., 1979. Orbits of ancient and medieval comets. {\it Publications of the Astronomical Society of  Japan}, 31, 257-270.
\vskip 2.0 em
Hasegawa, I., 1980. Catalogue of ancient and naked-eye comets. {\it Vistas in Astronomy}, 24, 59-102.
\vskip 2.0 em
Hiltebeitel, A., 1977. Nahu$\d {s}$a in the Skies: A Human King of Heaven. {\it History of Religions}, 16, 329-350.
\vskip 2.0 em
Ho, P.Y., 1962. Ancient and mediaeval observations of comets and novae in Chinese sources. {\it Vistas in Astronomy}, 5, 127-225.
\vskip 2.0 em
Hughes, D. W., 1985. The position of earth at previous apparitions of Halley's comet. {\it Quarterly Journal of the Royal Astronomical Society}, 26,   513-520.
\vskip 2.0 em
Iyengar, R. N., 2006. On Some Comet Observations in Ancient India. {\it Journal of Geological Society of India}, 67, 289-294.
\vskip 2.0 em
Iyengar, R. N., 2008. Archaic Astronomy of Parasara and Vrddha Garga. {\it Indian Journal of History of Science}, 43,  1-27.
\vskip 2.0 em
Iyengar, R. N., 2010. Comets and Meteoritic Showers in the Rgveda and their significance. {\it Indian Journal of History of Science}, 45, 1-32.
\vskip 2.0 em
Kanda, S., 1935. {\it Astronomical Materials in Japanese History}. Tokyo, Maruzen (in Japanese).
\vskip 2.0 em
Kanda, S., 1947. {\it Astronomical and Meteorological Materials in Japanese History}. Tokyo, Ashikabi (in Japanese).
\vskip 2.0 em
Kane, P. V., 1975. {\it History of Dharmasastra.} Volume 5. Poona,  Bhandarkar Oriental Research Institute.
\vskip 2.0 em
Kiang, T., 1972. The Past Orbit of Halley's Comet. {\it Memoirs of the Royal Astronomical Society}, 76, 27-66.
\vskip 2.0 em
Kochhar, R.,  2010. R$\bar{\mbox{{a}}}$hu and Ketu in Mythological and Astronomological Contexts. {\it Indian Journal of History of Science}, 45, 287-297.
\vskip 2.0 em
Kronk, G.W., 2003. {\it Cometography: A Catalog of Comets}. Volume 2: 1800-1899. Cambridge, Cambridge University Press.
\vskip 2.0 em
Mahadevan, I. 1986.  Agastya Legend and the Indus Civilization. {\it Journal of Tamil Studies}, 30, 24-37.
\vskip 2.0 em
Miki, M. and Yano, M.,  2010. A Study on the Atharvaveda-Parisista 50-57 with Special Reference to the Kurmavibhaga. {\it Journal of Indian and Buddhist Studies}, 58, 1126-1133.
\vskip 2.0 em
Parpola, A., 2009. ‘Hind Leg’+‘Fish’: Towards Further Understanding of the Indus Script. {\it Scripta}, 1,  37-76.
\vskip 2.0 em
Parpola, A., 2010. A Dravidian solution to the Indus script problem. A paper presented at the World Classical Tamil Conference, Coimbatore.  Chennai, Central Institute of Classical Tamil.
\vskip 2.0 em
Pingre, A.G., 1783. {\it Cometographie}. Paris, Imprimerie Royale.
\vskip 2.0 em
Pingree, D. E., 1981. {\it Jyotihsastra: Astral and Mathematical Literature}. Wiesbaden: Otto Harrassowitz.
\vskip 2.0 em
Robertson, W., 1831. {\it Philosophical Transactions of the Royal Society of London}, 121, 1-8.
Sachs, A., 1974. Babylonian observational astronomy. {\it Philosophical Transactions of the Royal Society of London A: Mathematical, Physical and Engineering Sciences}, 276(1257), 43-50.
\vskip 2.0 em
Sekiguchi, R., 1917a. Ancient Records of Comets in Korea. {\it Josen Kodai Kansoku Kiroku Chosa Hokoku}, 177-194  (in Japanese).
\vskip 2.0 em
Sekiguchi, R., 1917b. Meteor showers in Korean Records. {\it Josen Kodai Kansoku Kiroku Chosa Hokoku}, 196-200  (in Japanese).
\vskip 2.0 em
Sharma, S. D.  1987. Periodic Nature of Cometary Motions as Known to Indian Astronomers before Eleventh Century A.D. {\it History  of Oriental Astronomy, IAU Colloquium 91}. Cambridge, Cambridge University  Press. Pp.109-112.
\vskip 2.0 em
Stephenson, F.R., 1988. November. Oriental Star Maps. In Debarbat, S. (ed.). {\it Proceedings of the 133-rd  Symposium of the International Astronomical Union}. Cambridge, Cambridge University Press.  Pp.11-22.

\vskip 2.0 em
Stephenson, F.R. and Yau, K.K.C., 1985. Far Eastern observations of Halley's comet-240 BC to AD 
1368. {\it Journal of the British Interplanetary Society}, 38, 195-216.
\vskip 2.0 em
Stephenson, F.R., Yau, K.K.C. and Hunger, H., 1985. Records of Halley's comet on Babylonian tablets. {\it Nature}, 314, 587-592.
\vskip 2.0 em
Subbarayappa, B. V.  and Sarma, K. V., 1985. {\it Indian Astronomy - A source-book}. Bombay, Nehru Centre.
\vskip 2.0 em
Tsu, W.S., 1934. The observations of Halley's comet in Chinese history. {\it Popular Astronomy}, 42, 191-201.
\vskip 2.0 em
Vahia, M. N.  and Subbarayappa, B. V., 2011. Eclipses in ancient India.  In  Soma, M.  and  Tanikawa, K. (eds.). {\it Proceedings of the
Fourth Symposium on History of Astronomy}. Tokyo, National Astronomical Observatory of Japan. Pp.16-19.
\vskip 2.0 em
Williams, J., 2014. {\it Observations of Comets from BC 611 to AD 1640}. Cambridge, Cambridge University Press.
\vskip 2.0 em
Witzel, M., 1995. Early Indian History: Linguistic and Textual Parameters. In  Erdosy, G. (ed.).  {\it Indian Philology and South Asian Studies}. Berlin, Walter de Gruyter \& Co. Pp.85-125.
\vskip 2.0 em
Xu, Z., Pankenier, W. and Jiang, Y., 2000. {\it East-Asian Archaeoastronomy: Historical Records of Astronomical Observations of China, Japan and Korea}. Singapore, Gordon and Breach Science Publishers.
\vskip 2.0 em
Yano, M. 2003. Calendar, Astrology and Astronomy. In Flood, G. (ed.). {\it Blackwell Companion to Hinduism}. London, Blackwell Publishing. Pp. 376 - 392
\vskip 2.0 em
Yeomans, D.K. and Kiang, T., 1981. The long-term motion of comet Halley. {\it Monthly Notices of the Royal Astronomical Society}, 197, 633-646.
\vskip 2.0 em
Yeomans, D.K., Rahe, J. and Freitag, R.S., 1986. The history of comet Halley. {\it Journal of the Royal Astronomical Society of Canada}, 80, 62-86.

%
%
%
%
\end{document}